\documentclass[a4paper,11pt]{article}
\usepackage[utf8x]{inputenc}
\usepackage{graphicx}
\usepackage{amsmath}
\usepackage{hyperref}
\usepackage{commath}
\title{\textbf{Simple Harmonic Motion:\\Geometrical Solutions of Equations of Motions}}
\author{\textbf{Sridip Pal\footnote{sridippaliiser@gmail.com} , Soubhik Kumar\footnote{09soubhik@gmail.com}}\\Department of Physical Sciences,IISER-Kolkata\\}

\begin{document}

\maketitle\newpage

\begin{abstract}
Conventionally while we talk about geometry associated with a 
simple harmonic oscillator(hereafter denoted by,``SHO'') we draw a 
circle with a radius equal to the amplitude of SHO and imagine a particle 
moving along the perimeter with a frequency same as that of SHO and then the
x coordinate of the position of the particle gives the value of the displacement
of the SHO. Here we discuss another kind of diagram (polar plot) depicting SHO and try to get some idea about boundary 
and initial value problem.\ref{ref1}
\end{abstract}

\section{Introduction}
\hspace{15pt}SHO's are ubiquitous in physics.Any potential admitting stable equilibrium can
be expanded as a Taylor Series around the point of equilibrium and can be shown
to have a Simple Harmonic Oscillatory behavior for small oscillation around
that point.On the other hand, sometimes geometry
helps us visualize certain things compared to the algebra involved in the problem.
The equation of motion governing the motion of SHO is:
\begin{equation}
\label{equation1}
 \frac{d^{2}r}{dt^{2}}=-\omega^{2}r
\end{equation}
where $\omega$ is the natural frequency of the oscillator,$t$ is time of observation and $r$ is displacement from origin i.e point of stable equilibrium.
The solution to this equation is,
\begin{center}
\begin{equation}
 \label{re}
r=a\cos(\omega t-\phi)
\end{equation}
\end{center}
where $a$ and $\phi$ are respectively the amplitude and the initial phase, 
determined from given initial or boundary conditions.\\ First, justification is given why
the diagram in \figref{Fig:1}, what we will be dealing with, depicts SHO. Then, using this diagram, 
we solve a boundary value problem in which position of the SHO is specified to be $x_{1}$,$x_{2}$ at time instances 
respectively $t_{1},t_{2}$, and we are required to find the amplitude and intial phase of the oscillator.We also do the 
same for an initial value problem, where at one instant of time, position and velocity of the oscillator is given.\\
Of course the boundary value problem can be solved algrebaically in a straight forward manner.
The geometry supplies us with a new way of obtaining the familiar results. 

\section{The Diagram}
\begin{figure}[ht!]
\centering
\includegraphics[scale=0.7,keepaspectratio=true]{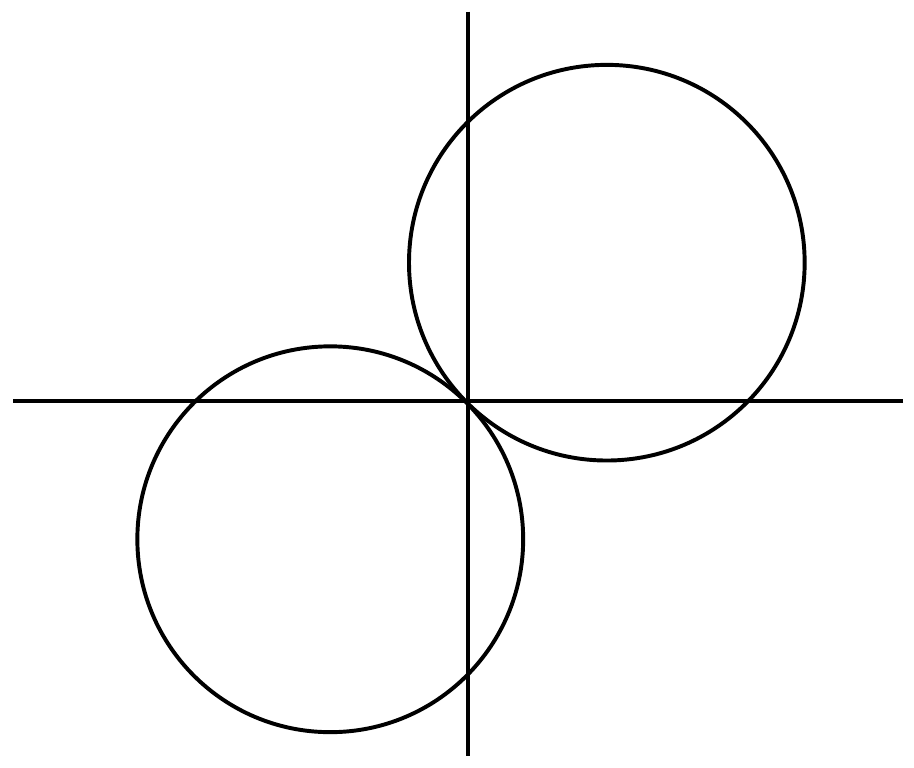}
\caption{The Solution Circles}
\label{Fig:1}
\end{figure}

\hspace{15pt} 
In \figref{Fig:1}, two circles, of same radius and touching each other externally, are drawn. Their radius and orientations are to be specified later.
We claim that the following \figref{Fig:1} represents motion of a SHO in the sense that if a point is moving along this curve then its distance from the origin vary with time in the same way as the distance(not, displacement) of a SHO from the origin. We also identify the polar angle the point makes with X-axis to be $\omega t$

\subsection{Proof of the Claim}
\begin{figure}[!ht] 
\centering
 \includegraphics[scale=1,keepaspectratio=true]{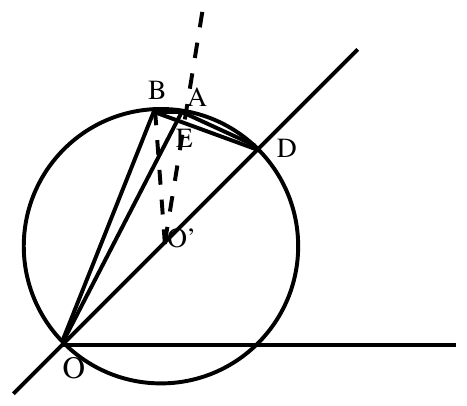}
 \caption{Geometric Solution to Equation of Motion}
 \label{Fig:2}
\end{figure}

Note,to prove the claim we only need one circle.[see \figref{Fig:2}]
Let $O^{\prime}$ be the centre of the circle, with $OD$ being the diameter
 whose length is $a$. Draw,a chord $OA$. Let, $OA$ be $r$. Suppose $OA$ makes an angle 
$\theta$ with the diameter.So, $OA = a \cos(\theta)$ since,$\angle OAD$ is $\pi /2$. Since, 
$r = a \cos \theta$, so \eqref{equation1} is satisfied. And thus our claim is proved.
\section{Boundary Value Problem} 
\hspace{15pt}Now we come to the boundary value problem. Suppose, we are given 
positions of the oscillators at two time instants,
$x(t_{1})=x_{1}$ and $x(t_{2})=x_{2}$ i.e. in a diagram like \figref{Fig:3}, two lines from the origin $OA$ 
and $OB$ having lengths $|x_{1}|$ and $|x_{2}|$ respectively, are given(\figref{Fig:3}). So as per previous convention,
$\angle AOB  = \omega(t_{2}-t_{1})$.Now,
 \begin{enumerate}
\item If $\angle AOB <\pi$(in anticlockwise sense), then draw the circumcircle of $\triangle AOB$.
 \item If $\angle AOB >\pi$(in anticlockwise sense),then reflect $OB$ through the origin and define the reflected line as $OB$, now we have made sure that $\angle AOB <\pi$, in fact $\angle AOB =\omega(t_{2}-t_{1})-\pi$,now we can draw the circumcircle of $\triangle AOB$.\footnote{The case $\angle AOB  = \pi$ is discussed later.}
 \end{enumerate}
\subsection{Finding the Amlitude and Initial Phase}
Reflect this circle about origin to get a pair of externally touching
circles and note $OD^{\prime}$ and $OB^{\prime}$ are the reflections of $OD$ and $OB$ respectively.Let $\angle AOB=\alpha$.We note that the diameter,$OD$ of the circle is amplitude; and the angle it makes with the X axis is minus of the initial phase $\phi$.
\begin{figure}[!ht]
\centering
 \includegraphics[scale=0.80,keepaspectratio=True]{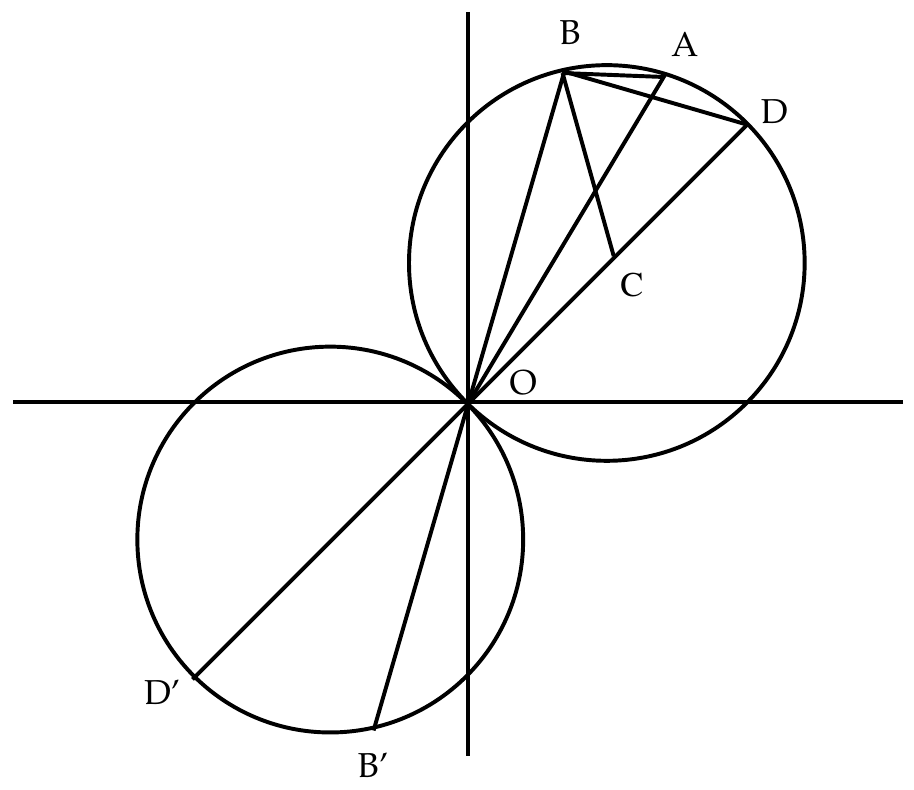}
\caption{Boundary Value Problem}
\label{Fig:3}
\end{figure}
First assume both of $x_{1}$ and $x_{2}$ are positive and then $\alpha = \omega(t_{2}-t_{1})$.Later we relax this restriction. Join points to get $AB$, $BD$.Since diameter is the biggest of all the chords, so
the diameter $OD$ is the amplitude of oscillation. We note that,
\begin{equation}
 \angle BDO= \angle BAO
\end{equation} 
since the angles are circumferential and generated by the same chord $OB$. Applying $\sin$ rule[\ref{ref2}] on$ \triangle AOB$,[\figref{Fig:3}]
\begin{equation}
\frac{AB}{\sin\alpha}=\frac{OB}{\sin\angle BAO}.
\end{equation} 
Combining we get,
\begin{equation}
\label{y}
\frac{AB}{\sin\alpha}=\frac{OB}{\sin\angle BDO}. 
\end{equation}
As $OD$ is
diameter so,$\angle OBD=\pi/2$,and $\sin$ rule[\ref{ref2}] on $\triangle ODB $ implies:
\begin{eqnarray}
\label{5}
 \frac{OB}{\sin\angle BDO}=OD\\
\Rightarrow \frac{AB}{\sin\alpha}=OD
\end{eqnarray}
where the last equation follows using \eqref{y}.
From the triangle $\triangle AOB$ using cosine rule[\ref{ref2}] we get:
\begin{equation}
\label{masterequation1}
 AB^{2}=x_{1}^{2}+x_{2}^{2}-2x_{1}x_{2}\cos\alpha
\end{equation}

From there we come up with:
\begin{equation}
\label{masterequation2}
 OD^{2}=\frac{x_{1}^{2}+x_{2}^{2}-2x_{1}x_{2}\cos\alpha}{\sin^{2}\alpha}
\end{equation}

$OD$ gives the amplitude of the SHO. Now, if $x_{1}$ and $x_{2}$ have opposite signs then $\omega(t_{2}-t_{1}) > \pi$ and 
a reflection of $OB$ is required, and then $\alpha=\omega(t_{2}-t_{1})-\pi$;so altogether the form of \eqref{masterequation1},
\eqref{masterequation2} is retained if we redifine $\alpha  = \omega(t_{2}-t_{1})-\pi$. And lastly if both of $x_{1}$ and 
$x_{2}$ are negative the form is same as \eqref{masterequation1},
\eqref{masterequation2} without any sort of redefinition of $\alpha$.\\
So in general, amplitude $a$ is,
\begin{equation}
\label{9}
a = \sqrt{\frac{x_{1}^{2}+x_{2}^{2}-2x_{1}x_{2}\cos(\omega(t_{2}-t_{1}))}{\sin^{2}(\omega(t_{2}-t_{1}))}}
\end{equation}
To know $\phi$ we are required to measure the angle, diameter makes with the $X$
axis. From the figure it is clear:
\begin{equation}
 \phi=\omega t_{1}-\arccos\frac{x_{1}}{a}
\end{equation}
So we have the complete solution with amplitude and initial phase.
\subsection{Uniqueness of Solution}
The uniqueness of the solution lies in the fact whether you can draw those
circles uniquely or not, given the boundary conditions.
\subsubsection{Case:1}
\begin{equation}
 \omega(t_{2}-t_{1})=2n\pi
\end{equation}
\hspace{15pt}It means, we are specifying the postion after integral time period.So for $x_{1}\neq x_{2}$ we do not have
any solution at all.If they are equal,it means $OA$ and $OB$ are identical,and the problem can not admit an unique solution as we require at least three
non collinear points to construct a circle uniquely. With $2$ points (including origin) we
can have infinitely many circles each corresponding to a solution satisfying given
boundary condition.
\subsubsection{Case:2}
\begin{equation}
 \omega(t_{2}-t_{1})=(2n+1)\pi
\end{equation}

\hspace{15pt}If the boundary condition does not satisfy $x_{1}=-x_{2}$, the system does not admit any
solution. Else $OA$ and $OB$ are diametrically opposite, and $\angle AOB = \pi$.Here also we can not draw any unique externally touching circle as the three points including origin is collinear. So 
we can have infinitely many circles each corresponding to a solution satisfying
boundary condition.
\subsubsection{Case:3}
\label{case3} 
\begin{equation}
x_{1}\ or\ x_{2}=0
\end{equation}
\hspace{15pt}In that case we do not have three non collinear points.But we can still come up
with an unique solution(satisfying boundary condition)
by drawing a circle with diameter of \textbf{nonzero} $\frac{x_{i}}{\cos(\omega t_{i}-\phi)}$ and then reflecting it
about origin.
\subsubsection{Case:4}
\label{case4}
\begin{equation}
 x_{1}\ and\ x_{2}=0
\end{equation}
\hspace{15pt}We have infinitely many solution provided $\omega(t_{2}-t_{1})=n\pi$,with one
solution corresponding to the situation that the particle does not oscillate at all.
\section{Initial Value Problem}
\begin{figure}[!ht]
 \centering
 \includegraphics{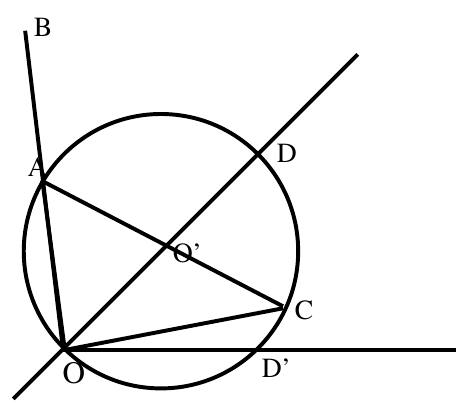}
 \caption{Initial Value Problem}
 \label{Fig:4}
\end{figure}

\hspace{15pt}First of all, note the solution to SHO \eqref{re} implies, $\frac{dr}{d\theta}=-a\sin\theta $ where
$\theta=\omega t-\phi$.Geometrically,what it means is the velocity vector can be obtained by
rotating the radius vector by $\pi/2$ in countercloclkwise sense and multiplying it by $\omega$.
In case of initial value problem we are given position and velocity at one
instant of time, so if we divide the magnitude of velocity vector
by $\omega$ we essentially obtain position vector after a quarter time period i.e $\pi/2$,
so our problem reduces to a boundary value problem.
The idea is stated below:

Let us consider the following \figref{Fig:4} with $OA=x(0)=x_{1}$ and $OB=v(0)=v_{1}$, where $x_{1}$ and
$v_{1}$ are initial displacement and initial velocity of the SHO.Rotate $OB$
clockwise by an angle of $\frac{\pi}{2}$.Cut 
$OC$ such that $\omega \times OC=OB$.Now draw the circumcircle of $\triangle AOC$.The
diameter of the circle will give
amplitude of oscillation while the angle it makes with X axis will give the
initial phase $\phi$.
Since our problem is now reduced to a boundary value problem as we previously
dealt with,only inital value problem fixes $\alpha$ ($\angle{AOC}$)[\figref{Fig:4}] equals to $\pi$/2.
\begin{eqnarray}
 OD^{2}=\frac{x_{1}^{2}+\frac{v_{1}^{2}}{\omega^{2}}}{\sin^{2}\frac{\pi}{2}}\\
OD^{2}=x_{1}^{2}+\frac{v_{1}^{2}}{\omega^{2}}
\end{eqnarray}

So the solution is:
\begin{equation}
 r(t)=OD\cos(\omega t-\arccos\frac{x_{1}}{OD})
\end{equation}
\subsection{Uniqueness of Solution}
\subsubsection{Case:1}
\begin{equation}
 x(0)\ or\ v(0)=0
\end{equation}
In the same way as discussed in case:\ref{case3} of boundary value (B.V) problem, we draw a unique circle of
diameter of $x(0)\ or\ \frac{v(0)}{\omega}$ whichever
is nonzero.
\subsubsection{Case:2}
\begin{equation}
 x(0)\ and\ v(0)=0
\end{equation}
Quite contrary to the case:\ref{case4} of the B.V problem here we have only one solution that
the particle does not oscillate at all because
in whatever way we draw the circle we have to traverse an angle of $n\pi$[see the \figref{Fig:4}] to come
to $0$,starting from $0$. But we know,the angle between $OA$ and $OC$
is $\frac{\pi}{2}$.So it can only be possible when the circles degenerate into a
point of origin i.e no oscillation.
\section{Generalization}
Although the above argument was entirely for SHO, but one can generalize it to any symmetric potential where oscillation is taking place about the point on symmetry axis. For those cases, instead of circles in \figref{Fig:1}, one gets other geometrical shapes. But one fact remains true, that the same two lobes are obtained, and as a whole the figure remains symmetric about $ y=x $ line. Now, because of this, if the values after half a time period are specified then, one can reflect the point about the origin, and then the two points coincide. So, we can be sure that no unique solution to the problem is possible, since with just origin and one point one draw any curves. Now, from Newton's law, we know we require just two points to draw a unique curve (except for special initial or boundary values), but at this moment it is not clear how can we argue about this necessity purely from geometry, for circles it was possible since we need just three points in general to draw a curve.
\section{Conclusion}
\hspace{15pt}It is satisfying that simple geometrical considerations solves the aforementioned problems. 
One usually uses algebraic methods to tackle these problems, but the
 uniqueness and existence of solutions is very transparent in this approach.These 
type of diagrams can give us an intuitive feeling about the system without any direct dependence on
algebric machinery.The whole scenario hinges on the fact \eqref{re} gives a diagram like mentioned above when plotted in ploar
co-ordinates.So any system (like Damped,Driven Harmonic Oscillator) whose solution can be reduced to this form \eqref{re} by proper substitution,we can employ this geometric
method to yield solution to boundary and initial value problems.

\section{Acknowledgements}
 \hspace{15pt}The authors would like to acknowledge a debt of gratitude to their teacher
 Dr. Ananda Dasgupta (Dept. of Physical Sciences, IISER-Kolkata) and
 Prof. S. Sengupta, Dept. of Theoretical Physics, IACS for their valuable
suggestions and comments. Thanks are also due to his father,Mr.Samir Kumar on behalf of
Soubhik Kumar, and Mr.Samir Ghosh on behalf of Sridip Pal to introduce them to
the fascinating world of geometry in school.

\section*{Reference}
\begin{enumerate}
\item \label{ref1}Mechanics, Berkeley Physics Course-Volume 1, Second edition (Special Indian Edition), Third reprint 2008, Tata McGraw-Hill
\item \label{ref2}Hall and Stevens, A School Geometry, Metric (Indian) Edition, Reprinted 2005, Radha Publishing House, Calcutta.
\end{enumerate}

\end{document}